\documentclass{pasj00}

\begin{document}
\SetRunningHead{Grechnev et al.}{Microwave negative bursts and
solar eruptions}

\title{Microwave Negative Bursts as Indications of Reconnection
between Eruptive Filaments and Large-Scale Coronal Magnetic
Environment}

\author{Victor \textsc{Grechnev}\altaffilmark{1},
 Irina \textsc{Kuzmenko}\altaffilmark{2},
 Arkadiy \textsc{Uralov}\altaffilmark{1},
 Ilya \textsc{Chertok}\altaffilmark{3},
 Alexey \textsc{Kochanov}\altaffilmark{1}}
 \altaffiltext{1}{Institute of
Solar-Terrestrial Physics SB RAS, Lermontov St. 126A, Irkutsk
664033, Russia}\email{grechnev@iszf.irk.ru}
\altaffiltext{2}{Ussuriysk Astrophysical Observatory, Solnechnaya
St. 21, Primorsky Krai, Gornotaezhnoe 692533,
Russia}\email{kuzmenko\_irina@mail.ru}
 \altaffiltext{3}{Pushkov Institute of Terrestrial Magnetism,
Ionosphere and Radio Wave Propagation (IZMIRAN), Troitsk, Moscow
Region, 142190 Russia}\email{ichertok@izmiran.ru}


\KeyWords{filament eruptions; microwave negative bursts, Sun:
coronal mass ejections (CMEs), Sun: radio radiation}

\maketitle

\begin{abstract}

Low-temperature plasma ejected in solar eruptions can screen
active regions as well as quiet solar areas. Absorption phenomena
can be observed in microwaves as `negative bursts' and in
different spectral domains. We analyze two very different recent
events with such phenomena and present an updated systematic view
of solar events associated with negative bursts. Related filament
eruptions can be normal, without essential changes of shape and
magnetic configuration, and `anomalous'. The latter are
characterized by disintegration of an eruptive filament and
dispersal of its remnants as a cloud over a large part of solar
disk. Such phenomena can be observed as giant depressions in the
He{\sc ii} 304~\AA\ line. One of possible scenarios for an
anomalous eruption is proposed in terms of reconnection of
filament's internal magnetic fields with external large-scale
coronal surrounding.

\end{abstract}


\section{Introduction}

Depressions of the total microwave flux discovered by
\citet{Covington1953} and called `negative bursts' were
interpreted by absorption in ejected filament material. Later
studies (e.g., \cite{Tanaka1960}) led to a general scenario of
screening of a compact source \citep{Sawyer1977}. The presumed
association of microwave negative bursts with solar eruptions
provides a basis for expectations to obtain additional information
about eruptions and their parameters.

A multi-spectral study of the 2004 July 13 solar event has led
\citet{Grechnev2008} to an unexpected conclusion about an
anomalous course of the filament eruption in this event. Unlike a
typical situation, the eruptive filament after the initial rise
apparently disintegrated into a large cloud of fragments, which
dispersed over almost a quarter of the visible solar disk, and
then mostly descended on the solar surface far from the eruption
site. The dispersed filament material screened the background
solar emission. The 2004 July 13 eruptive event has shown that in
addition to screening compact sources important can be screening
large quiet Sun's areas. The absorption phenomena were manifest in
a microwave negative burst being especially pronounced in a He{\sc
ii} 304~\AA\ image of SOHO/EIT \citep{Delab1995} as a large
darkening mismatching CME-related dimming observed in coronal
extreme-ultraviolet (EUV) bands. Among all of EUV emission lines,
in which observations are carried out, the He{\sc ii} 304~\AA\
line is best suited for studies of prominences and filaments due
to its temperature sensitivity range (maximum at 80\,000 K).
SOHO/EIT produced He{\sc ii} 304~\AA\ images usually once in 6
hours.

We have not found in the literature descriptions of similar
phenomena, except for the only report on the 2003 November 18
event \citep{Slemzin2004,Grechnev2005}, of which we were aware
(V.G. and I.C. were among the co-authors). A negative burst could
not be observed in this event due to subsequent flaring. The event
was studied in detail later \citep{Grechnev2013_I}. A review of
\citet{Gilbert2007} does not mention such phenomena. A reason why
anomalous eruptions were not well known previously seems to be
clear. Observations of eruptions in the H$\alpha$ line are limited
by the Doppler shift, which rapidly removes absorbing features
from the filter passband even with moderate line-of-sight
velocities. Observations in the He{\sc ii} 304~\AA\ line were
infrequent in the past.

Proceeding from the results of \citet{Grechnev2008}, we searched
for i)~events with negative microwave bursts recorded in Nobeyama
and Ussuriysk, and ii)~events, in which large darkenings were
observed in 304~\AA\ images. One more event with anomalous
eruption, which occurred on 1998 April 29 around 17:30~UT, was
revealed from EIT data \citep{ChertokGrechnev2003} and analyzed by
\citet{Grechnev2011}.

The events with negative bursts recorded at several frequencies
were analyzed by using a simple model developed for estimating
parameters of ejected material. The model allows one to calculate
the spectrum of the total solar radio flux by considering
contributions from the chromosphere, a screen constituted by
material of an eruptive filament `inserted' into the corona at
some height, and coronal layers both between the chromosphere and
the screen, and between the screen and the observer
\citep{Grechnev2008}. By comparing the frequency distribution of
the microwave absorption depth in an observed negative burst with
an absorption spectrum simulated with the model, it is possible to
estimate the kinetic temperature of the screen, its optical
thickness at each frequency, the area, and the height above the
chromosphere. Actual fluxes of compact sources were measured from
NoRH images and used among other input parameters of the model in
the simulations.

The studies of
\authorcite{Grechnev2008} (\yearcite{Grechnev2008,Grechnev2011}) and
\citet{Kuzmenko2009} presented several events with negative bursts
detected mostly from measurements with NoRP and RT-2 (2.8 GHz) in
Ussuriysk \citep{Kuzmenko2008} and some events with a large
darkening in 304~\AA\ images. The estimated parameters are
consistent with the assumption about screening of the solar
emission by cool material of an eruptive filament. Among these
events, there were also those with anomalous eruptions.
\citet{Grechnev2008} concluded that magnetic reconnection was most
likely implicated in the observed transformation of the eruptive
filament, but a particular role of reconnection and a possible
scenario of such an anomalous eruption remained unclear.

Noticeable negative bursts were recently observed in two very
different events. The 2011 June 7 event was associated with a
powerful flare and strong impulsive microwave burst. The 2011
December 13 event has not produced any detectable enhancement in
total flux records. We firstly analyze the two recent events. Then
we present a possible scenario of an anomalous eruption and
discuss distinctive particularities and possible common properties
of such events. On the basis of these considerations and results
of previous studies we endeavor to supplement a systematic view of
solar eruptions associated with negative bursts.

\section{Recent Events with Negative Bursts}

Our analysis of events with microwave negative bursts uses total
flux measurements at several frequencies routinely carried out
with the following instruments:

\begin{itemize}

\item
 the Nobeyama Radio Polarimeters (NoRP,
\cite{Torii1979,Nakajima1985}); we use data at the frequencies of
1, 2, 3.75, 9.4, and 17 GHz;

\item
 the radiometers of the Learmont station of the USAF RSTN network;
we use data at the frequencies of 1.4, 2.7, 5.0, and 8.8 GHz;

\item
 the RT-2 radiometer observing at 2.8 GHz \citep{Kuzmenko2008}
operated by the Ussuriysk Astrophysical Observatory (station code
`VORO').

\end{itemize}

Parameters of quasi-stationary microwave sources were measured
from images produced by the Nobeyama Radioheliograph (NoRH,
\cite{Nakajima1994}) at a frequency of 17 GHz. Eruptive events
were analyzed by using EUV images produced by SOHO/EIT
\citep{Delab1995}, STEREO/EUVI \citep{Howard2008}, and SDO/AIA
\citep{Lemen2012AIA} as well as NoRH images as long as they were
available. We also used the images produced with the Siberian
Solar Radio Telescope at 5.7 GHz (SSRT,
\cite{Smolkov1986,Grechnev2003}).

\subsection{Event I: 2011 December 13}

\begin{figure*}
  \begin{center}
    \FigureFile(80mm,125mm){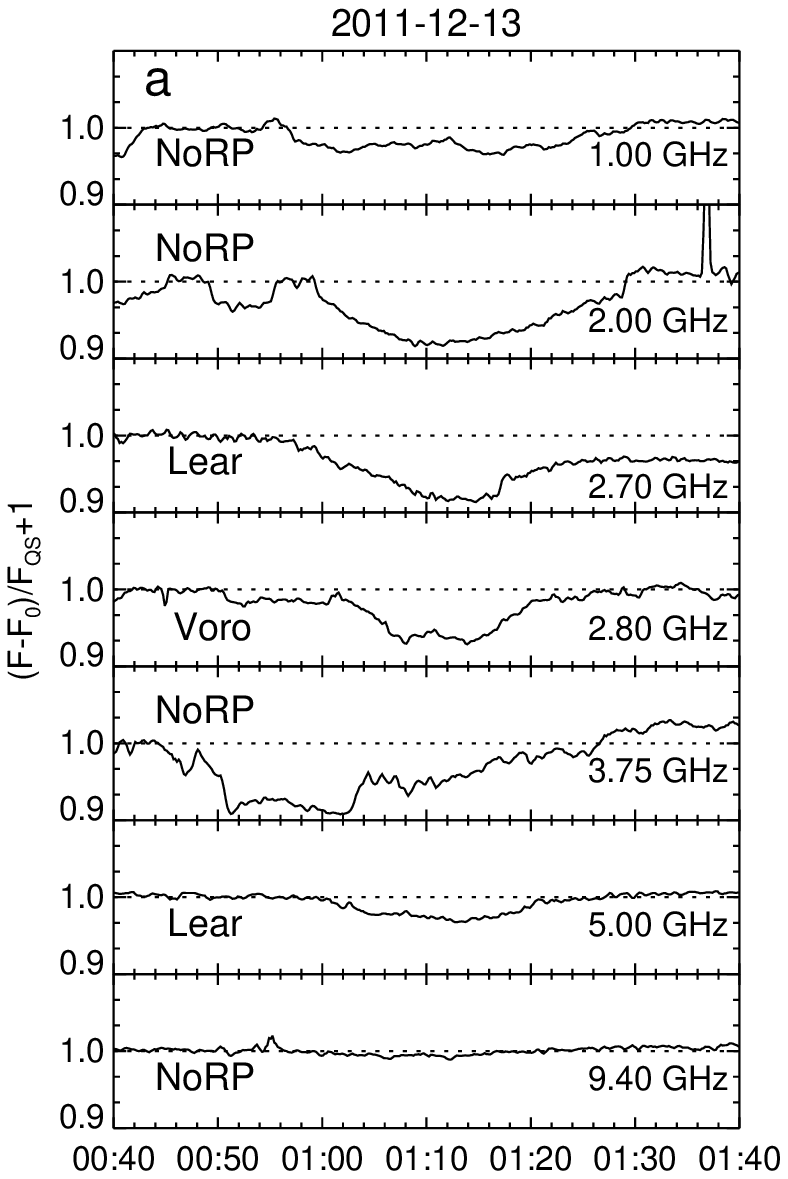}
    \FigureFile(80mm,125mm){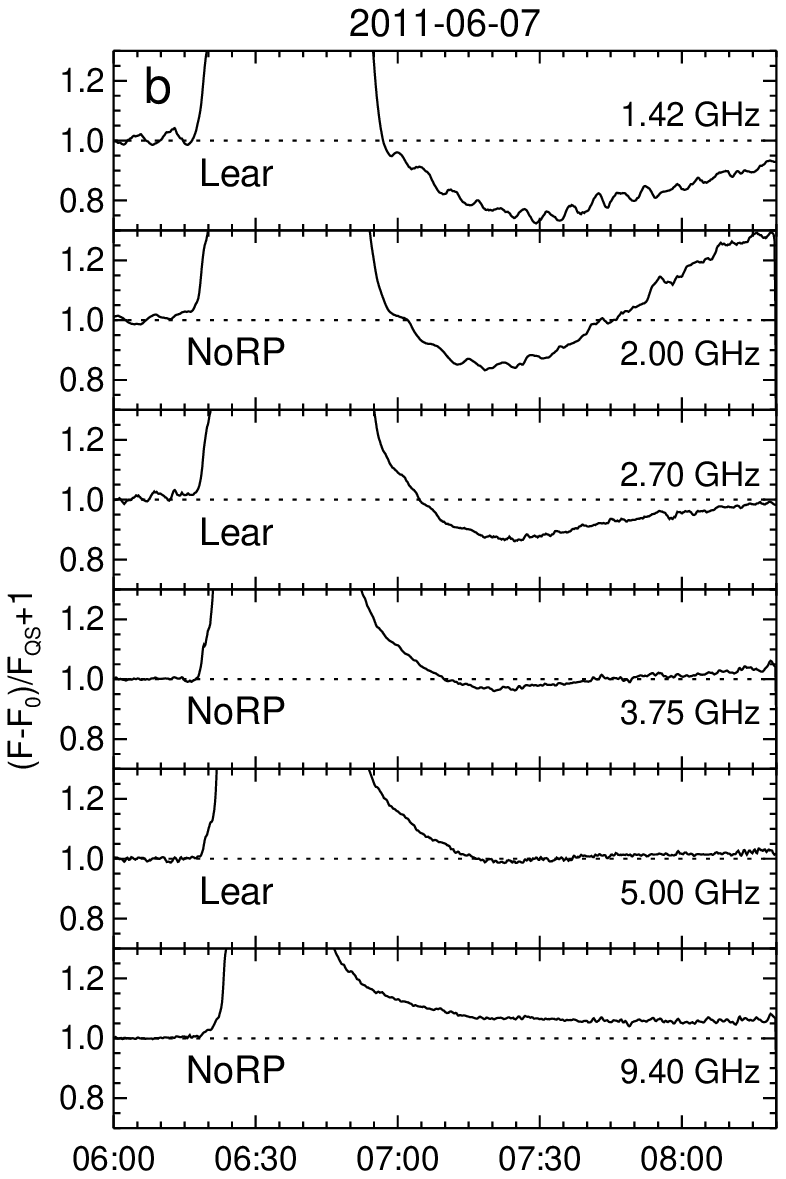}
  \end{center}
  \caption{Negative bursts recorded on 2011 December 13 (a)
and on 2011 June 7 (b). Time profiles are normalized to the quiet
Sun's levels at different frequencies.}
   \label{fig:timeprofs}
 \end{figure*}

\begin{figure*}
  \begin{center}
    \FigureFile(150mm,115mm){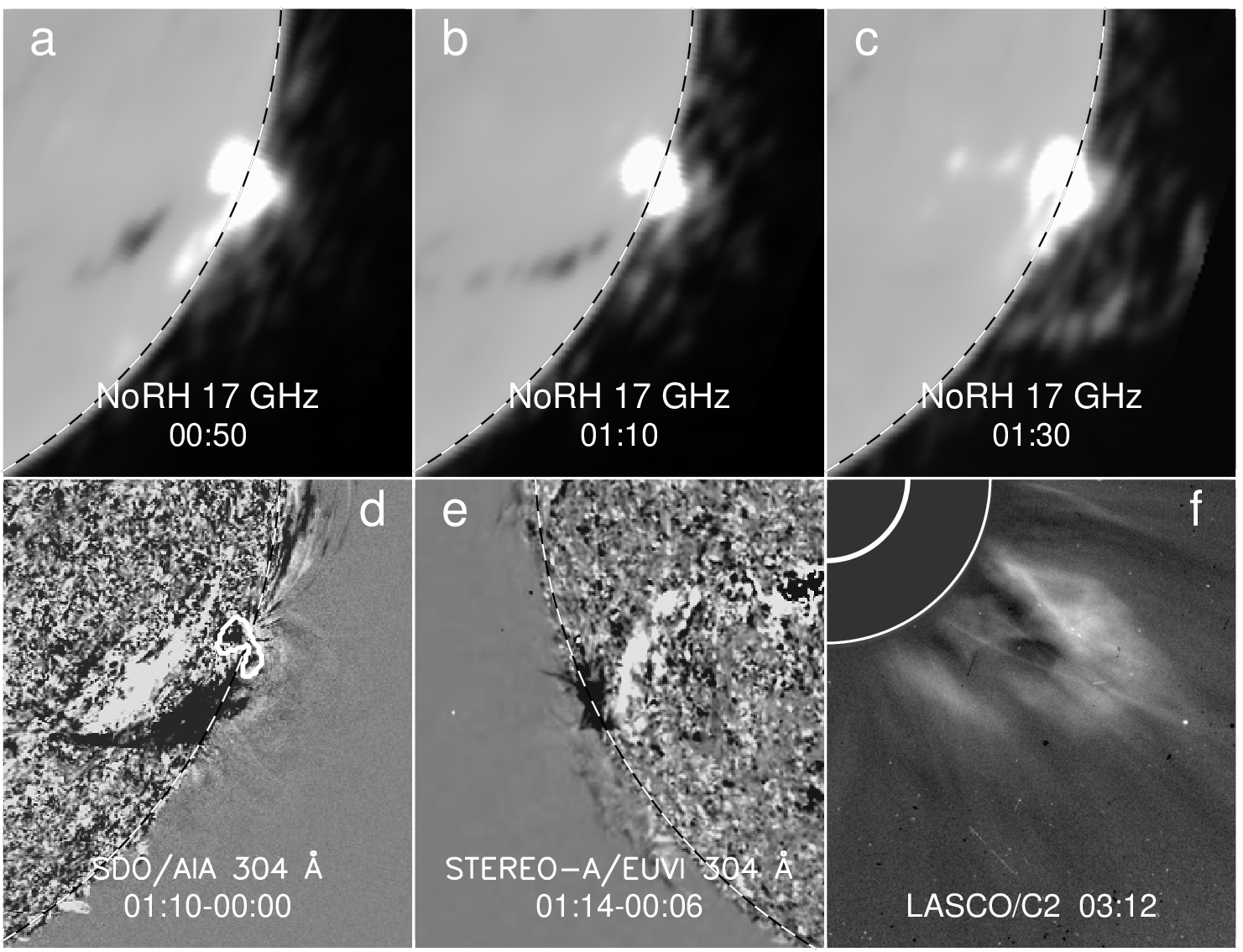}
  \end{center}
  \caption{Filament eruption on 2011 December 13: a--c)~microwave NoRH
images without subtraction, d)~SDO/AIA and e)~STEREO-A/EUVI
difference images at 304~\AA, e)~LASCO/C2 ratio image. The white
contours in panel (d) outline the screened sources in the NoRH
17~GHz image. The dashed circle denotes the solar limb. The white
solid circles in panel (f) denote the solar limb and the inner
boundary of the LASCO/C2 field of view.}
 \label{fig:20111213_images}
 \end{figure*}

On 2011 December 13, an isolated negative burst was observed.
Unlike a typical situation, it was not preceded by a flare burst.
A possible enhancement of the soft X-ray flux was inconspicuous in
GOES plots. Figure~\ref{fig:timeprofs}a presents the total flux
time profiles of the microwave emission with subtracted pre-burst
levels, normalized to the quiet Sun's total flux at each
frequency, and smoothed over 30~s. As
Figure~\ref{fig:20111213_images}a--d shows, depression of the
microwave emission was caused by screening remote compact sources
by a steadily expanding quiescent filament. The eruptive filament
was observed as a moving dark feature covering the remote sources
on the limb both in NoRH 17~GHz and SDO/AIA 304~\AA\ images. The
fluxes of the screened compact sources estimated from the 17~GHz
NoRH image at 01:10 were $\approx 4$ sfu. The positions of the
screened sources are shown by the white contours in
Figure~\ref{fig:20111213_images}d.

The difference SDO/AIA 304~\AA\ image in in
Figure~\ref{fig:20111213_images}d shows at 01:10 the area of the
darkening at a level of $-20\%$ of the brightness to be 0.5\% of
the solar disk. The maximum depth of the depression reached
$-54\%$. The eruptive filament was also observed from the vantage
point of STEREO-A $110^{\circ}$ west of the Earth. The
STEREO-A/EUVI 304~\AA\ image in Figure~\ref{fig:20111213_images}e
allows us to estimate the geometrical depth of the eruptive
filament $L \approx 100$ Mm in the direction close to the line of
sight of an observer on Earth.

With these quantities, parameters of the screen were estimated
from the model simulations of the absorption spectrum. The
estimated temperature of the filament material is 10\,000~K. The
estimated area is about twice larger than the 304~\AA\ image
shows. This is possible, because the body of the filament
conspicuous in Figure~\ref{fig:20111213_images}e can be surrounded
by a sheath invisible at 304~\AA\ but contributing to microwaves.
The estimated mass of the filament of $3 \times 10^{14}$~g is less
than those provided by estimations in other events (see
Table~\ref{tab:burst_list}), but still comparable with typical
masses of filaments.

Eruption of the filament far from active regions determined
weakness of the emission from the post-eruption arcade, although
the arcade was detectable, e.g., in 304~\AA\ images in
Figure~\ref{fig:20111213_images}d,e. Weakness of the arcade
emission determined the absence of a detectable burst in total
flux records before the negative burst. This weak eruptive event
has produced a CME visible in a SOHO/LASCO/C2 image in
Figure~\ref{fig:20111213_images}f. Properties of the 2011 December
13 event appear to be typical of non-flare-related eruptions
\citep{Chertok2009}. Among these properties is gradual
acceleration of the CME in LASCO/C2 images according to a
height-time plot in the SOHO LASCO CME Catalog
(http:/\negthinspace/cdaw.gsfc.nasa.gov/CME\_list,
\cite{Yashiro2004}). Finally we note that the 2011 December 13
eruption and negative burst resemble the event discussed by
\citet{Maksimov1991}, which also occurred according to the
`classical' screening scenario \citep{Sawyer1977}.

\subsection{Event II: 2011 June 7}

The 2011 June 7 event occurred in active region 11226 (S22~W64,
$\beta$-configuration). Images produced before the event in the
Big Bear Solar Observatory in the H$\alpha$ line show a small
filament in the active region. The pre-eruptive filament is also
visible in SDO/AIA 304~\AA\ images. The filament started to rise
by 06:11. The related flare of M2.5 GOES importance started at
06:16.

\begin{figure*}
  \begin{center}
    \FigureFile(170mm,56mm){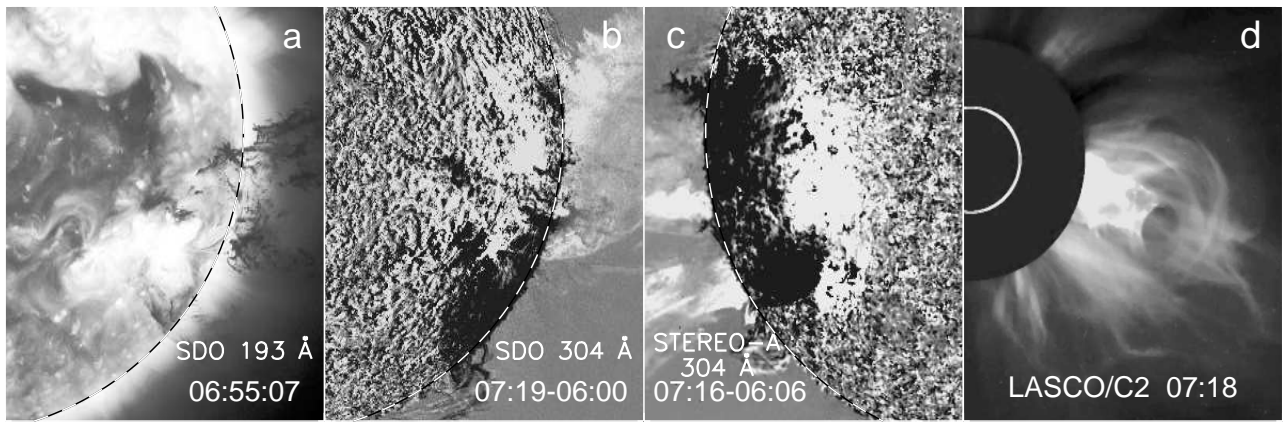}
  \end{center}
  \caption{2011 June 7 event: SDO/AIA image at 193~\AA\ (a),
SDO and STEREO-A differences at 304~\AA\ (b,c), SOHO/LASCO/C2
image of the CME (d). The dashed circle denotes the solar limb.}
 \label{fig:20110607_sdo_stereo}
 \end{figure*}

Importance of this event for our study is determined by an
anomalous eruption, which was the most demonstrative one ever
observed. The eruptive filament after the initial lift-off by
06:25 apparently disintegrated into a huge dome of jet-like
fragments, most of which turned back to the solar surface and
produced a `coronal rain', which went on four hours. Absorbing
fragments of the dispersed filament are well visible not only at
304~\AA, but even in non-subtracted images at 193~\AA\ in
Figure~\ref{fig:20110607_sdo_stereo}a and in other coronal lines
up to 94~\AA\ (see also 211~\AA\ images in
Figure~\ref{fig:20110607_211}).

The SDO/AIA 304~\AA\ difference image in
Figure~\ref{fig:20110607_sdo_stereo}b shows a large darkening,
while some portion of the area subjected to the coronal rain was
bright. The area of the darkening at a level of the brightness
decrease of $-25\%$ reached 6.5\% of the solar disk. The deepest
local depression reached $-90\%$. The large brightening area
(2.7\% of the solar disk) indicates the presence of hotter
material there. The total area occupied by manifestations of
dispersed filament material was 6.6\% from SDO/AIA images and 8\%
from the STEREO-A/EUVI 304~\AA\ image in
Figure~\ref{fig:20110607_sdo_stereo}c.

\begin{figure}
  \begin{center}
    \FigureFile(85mm,185mm){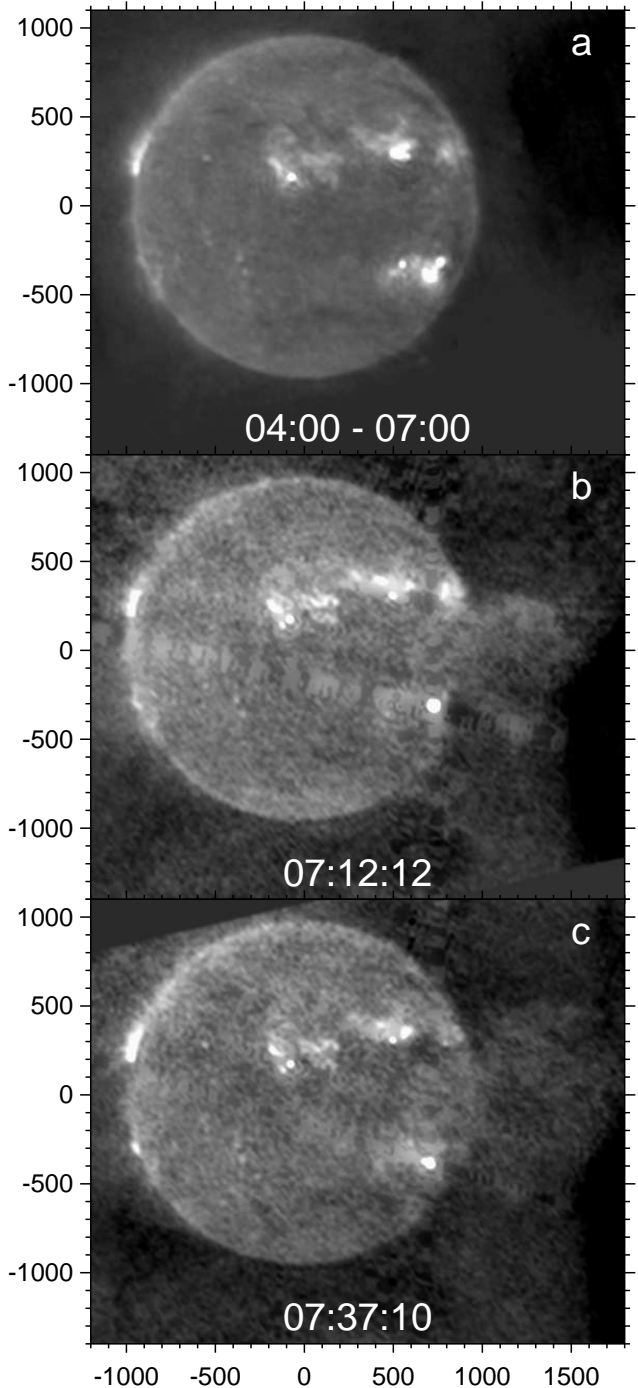}
  \end{center}
  \caption{SSRT observations on 2011 June 7 at 5.7 GHz: a)~pre-event image
averaged from 04:00 to 07:00; b,~c)~a huge cloud of dispersed
filament material expanding above the west limb. The axes show
hereafter arc seconds from the solar disk center.}
  \label{fig:20110607_ssrt}
 \end{figure}

NoRH recorded the 2011 June 7 event by the end of the observations
at 06:30. The negative burst began after 06:50. For this event
microwave images produced with SSRT at a frequency of 5.7~GHz are
available. Figure~\ref{fig:20110607_ssrt}a shows a pre-event SSRT
image averaged over several frames observed from 04:00 to 07:00.
Figures~\ref{fig:20110607_ssrt}b and \ref{fig:20110607_ssrt}c show
the Sun observed during the negative burst. Dark portions of a
disk in the right part of the images are due to adjacent negative
interference beams of the SSRT that limit the field of view. To
emphasize low-brightness features of interest, we had to limit
bright sources from above (their brightness temperatures exceed
$1.7 \times 10^5$~K) which produces impression of their
saturation.

The 5.7 GHz SSRT images in Figures~\ref{fig:20110607_ssrt}b and
\ref{fig:20110607_ssrt}c present a huge cloud consisting of
dispersed filament material above the limb and, possibly,
partially on the solar disk. Inevitable residual side lobes still
remaining after the CLEAN routine (long fringe-like stripes
crossing bright sources) somewhat distort the image. A gradual
expansion of the cloud is visible in the two lower panels.

The cloud is visible in microwaves undoubtedly due to thermal
free-free emission. The optical thickness of the cloud can be less
than in EUV so that the microwave brightness responds to emissions
from different depths averaged along the line of sight. If the
optical thickness is large, then the microwave brightness
temperature is equal to the kinetic temperature of a source.
Bright parts of the cloud are similar in brightness to the quiet
Sun, whose brightness temperature is 16\,000~K at 5.7 GHz. Thus,
bright portions of the cloud had an average temperature of $\gtsim
16\,000$~K, which suggests appreciable heating of dispersed
filament material. A reason for a darker appearance of the middle
part of the cloud is difficult to understand from a
single-frequency image. This can be due to either a lower
temperature or a lesser thickness.

This event is the only one of such a kind observed with SSRT so
far. Analysis of other events in which negative bursts were
observed has shown that the expected depression depth at the
relatively high frequency of 5.7 GHz was less than the sensitivity
of SSRT \citep{Grechnev2011}.

A negative burst was observed after the impulsive burst at
frequencies below 5~GHz. Figure~\ref{fig:timeprofs}b shows the
negative burst with the pre-burst levels subtracted from the time
profiles, normalized to the quiet Sun, and smoothed over 30~s like
Figure~\ref{fig:timeprofs}a. As observations and model simulations
show, the negative burst in this event was due to screening both a
compact source and a large quiet Sun's area. Since the negative
burst occurred after completion of NoRH observations, we use in
our modeling the flux of the screened compact radio source of
$\approx 3$ sfu measured from a NoRH 17 GHz image at 06:00. The
mass of absorbing material estimated from the radio spectrum is $6
\times 10^{15}$~g. The estimated average temperature is higher
than that in the preceding event, $3 \times 10^4$~K, which seems
to be consistent with the presence of large brightening areas in
Figure~\ref{fig:20110607_sdo_stereo}b,c. The temperature is
consistent with the estimate from the SSRT data. The area of the
absorbing cloud estimated from model simulations is almost twice
larger than that estimated from the 304~\AA\ images. Note,
however, that the model used in the estimations takes account of
the on-disk projection of the screening material and does not
consider an exotic possibility of a sporadic ``another Sun''
constituted by the huge microwave-emitting cloud, and therefore
the estimates for this event are less reliable.

The top part of the filament, which was initially dark, then
brightened up in the course of its lift-off, while trailing
material remained predominantly dark. Approximate estimates of
kinematics of the leading filament's edge have provided its
initial velocity (06:10--06:20) in the plane of the sky of about
17~km~s$^{-1}$. By 06:24 the filament accelerated up to
540~km~s$^{-1}$ and then started to decelerate. After 06:34 the
bright top of the eruption reached the edge of the field of view
of SDO/AIA that does not allow us to estimate its final velocity.
The last measurement provides 214~km~s$^{-1}$ for the leading
edge. The trailing dark portion also showed the highest speed
between 06:25 and 06:30 and decelerated later on.

The sharp filament eruption with strong acceleration followed by
deceleration must have produced a shock wave via the
impulsive-piston mechanism (see \cite{Grechnev2011_I}). This
expectation is confirmed by observation of an `EUV wave' in the
193~\AA\ channel starting from 06:24 and a type II radio burst
(Learmonth).

A fast CME was observed after the eruption. The CME in
Figure~\ref{fig:20110607_sdo_stereo}d resembles in shape the cross
section of a semi-torus with its straight poloidal axis being
close to the plane of the sky so that its circular toroidal axis
is close to the line of sight. Brighter features suggestive of
remnants of filament material are visible near the bottom of the
central tube of the torus (cf.
Figure~\ref{fig:scenario_anomalous}d,e,f). LASCO/C2 movies
available in the SOHO LASCO CME Catalog show at 09:30--11:30 that
some pieces of bright material reached heights of $(2-2.5)R_\odot$
above the solar surface and then fell back.

\section{Discussion}

\subsection{Revealed Events}

\begin{figure}
  \begin{center}
    \FigureFile(85mm,60mm){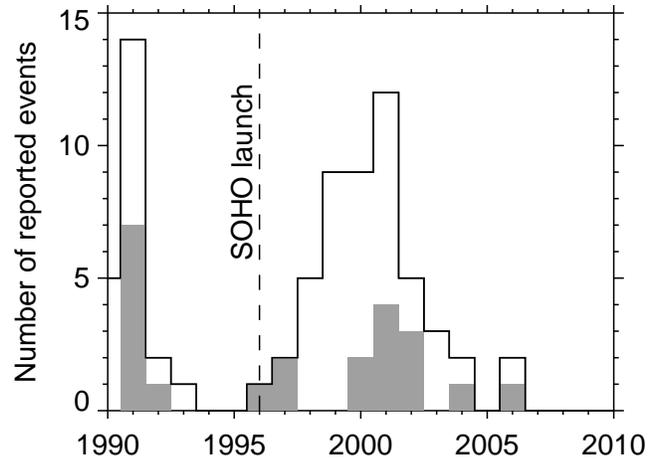}
  \end{center}
  \caption{Yearly occurrence of negative bursts in 1991--2010 according
to all reports in Solar-Geophysical Data (solid histogram) and
during the observational daytime in Nobeyama and Ussuriysk (shaded
bars). The vertical dashed line marks the date of SOHO launch.}
  \label{fig:statistics_neg_bursts}
 \end{figure}

Information about microwave negative bursts in the past years is
contained in reports on single-frequency observations of solar
radio bursts presented in the \textit{Solar-Geophysical Data}
bulletin issued from 1955 to 2009. Overview of reported events
with negative bursts (code `ABS') contained in
\textit{Solar-Geophysical Data} shows their very rare occurrence.
Figure~\ref{fig:statistics_neg_bursts} summarizes the yearly
number of the reports since 1991 (when observations in Ussuriysk
started) and up to 2009. Multiple reports on the same event have
been removed. The shaded histogram selects only those events which
could be observed in Nobeyama or Ussuriysk. The total number of
negative bursts reported in 1991--2009 within the observational
daytimes in Nobeyama and Ussuriysk was as small as 22. A detailed
analysis of such events has become possible when EUV observations
have become available, i.e., when SOHO was launched (the dashed
vertical line in Figure~\ref{fig:statistics_neg_bursts}). The rate
of yearly reported events mainly corresponds to the progression of
the solar cycle, which can be expected.

Table~\ref{tab:burst_list} lists characteristics of revealed
negative bursts (including two events considered in the preceding
section and two most recent events which have not yet been
analyzed) and estimated parameters of absorbing material. The
dates and times are related to the deepest depression. Column 4
presents a range of durations observed at different frequencies
(typically longer at lower frequencies). The depth in column 5
corresponds to a deepest depression expressed as a percentage of
the quiet Sun's total flux. Column 6 specifies a frequency at
which the deepest depression occurred. As column 10 of
Table~\ref{tab:burst_list} shows, more than 50\% of the revealed
event were not registered in \textit{Solar-Geophysical Data}. Even
if the missed events were included into the histogram in
Figure~\ref{fig:statistics_neg_bursts}, the total number of events
with negative bursts would remain small.

\begin{table*}
 \caption{Summary of revealed events with negative bursts}
 \label{tab:burst_list}
 \begin{center}
 \begin{tabular}{rlccrcrrrcc}
 \hline
\multicolumn{1}{c}{No.} & \multicolumn{1}{c}{Date} &
\multicolumn{1}{c}{Time} & \multicolumn{1}{c}{Duration} &
\multicolumn{1}{c}{Depth} & \multicolumn{1}{c}{$\nu$} & \multicolumn{1}{c}{$T_\mathrm{S}$} &
\multicolumn{1}{c}{$A_\mathrm{S}/A_{\odot}$} &
\multicolumn{1}{c}{M} & {SGD} & \multicolumn{1}{c}{Remarks}\\

\multicolumn{1}{c}{} & \multicolumn{1}{c}{} & {} & min &
\multicolumn{1}{c}{\%} & \multicolumn{1}{c}{GHz} & \multicolumn{1}{c}{K} &
\multicolumn{1}{c}{\%} &
\multicolumn{1}{c}{$10^{14}$~g} & {report} & {}\\

 \hline
\multicolumn{1}{c}{1} & \multicolumn{1}{c}{2} & \multicolumn{1}{c}{3} &
\multicolumn{1}{c}{4} & \multicolumn{1}{c}{5} & \multicolumn{1}{c}{6} &
\multicolumn{1}{c}{7} & \multicolumn{1}{c}{8} & \multicolumn{1}{c}{9} &
\multicolumn{1}{c}{10} & 11 \\
 \hline

1 & 1998-04-29 & 17:29 & 53\,--\,65 & 11 & 2.8 & {} & {} & {} & {Yes} & anomalous \\
2 & 2000-06-16 & 00:10 & 28\,--\,80 & 19 & 1.0 & 9000 & 6.0 & 13 & {No} & {} \\
3 & 2002-02-06 & 04:51 & 18\,--\,34 & 10 & 2.0 & 9000 & 3.5 & 20 & {Yes} & {} \\
4 & 2002-02-07 & 01:10 & 16\,--\,32 & 10 & 2.0 & 11000 & 4.0 & 15 & {No} & {} \\
5 & 2002-06-02 & 00:01 & 22\,--\,26 & 6 & 2.7 & 8000 & 2.0  & 9 & {Yes} & {}  \\
6 & 2003-05-28 & 00:10 & {10} & {} & {} & {} & {} & {} & {No} & {onset only} \\
7 & 2004-07-13 & 00:55 & 37\,--\,80 & 12 & 1\,--\,2.8 & 10000 & 6.4 & 15 & {No} & anomalous \\
8 & 2005-01-01 & 01:06 & 45\,--\,65 & 13 & 2.0 & 14000 & 5.0 & 20 & {No} & {} \\
9 & 2010-11-12 & 01:55 & {$\sim 25$} & 2 & 1.0 & {} & {} & {} & {--} & {} \\
10 & 2011-06-07 & 07:20 & 30\,--\,100 & 25 & 1.4 & 30000$^{*}$ & 15$^{*}$ & 60$^{*}$ & {--} & anomalous \\
11 & 2011-12-13 & 01:14 & 20\,--\,31 & 8 & 2.0 & 10000 & 1.2 & 3 & {--} & isolated \\
12 & 2012-01-12 & 01:26 &  37\,--\,56 & 6 & 2.0 & {} & {} & {} & {--} & {} \\
13 & 2013-02-06 & 00:43 & 15\,--\,45 & 12 & 2.0 & {} & {} & {} & {--} & {} \\

 \hline
 \end{tabular}
 \end{center}
 $^{*}$Estimated parameters are questionable due to
 exceptional characteristics of the ejected cloud
 \end{table*}


Analysis of several events has led \authorcite{Grechnev2008}
(\yearcite{Grechnev2008,Grechnev2011}) and \citet{Kuzmenko2009} to
the following conclusions. Screening both compact sources and
large quiet Sun's areas can be important. Negative bursts are best
observed at 1--5 GHz, where the corona is optically thin and
contrast of ejecta against the quiet Sun is sufficient.
Nevertheless, absorption sometimes appears at 17 GHz. A post-burst
decrease can be observed if a preceding flare is short and no
bursts occur afterwards. When negative bursts occur, large
darkening can appear at 304~\AA\ mismatching CME-related dimming
in coronal bands.

In all events screening of both compact sources and considerable
quiet Sun's areas occurred. Almost all negative bursts in the
analyzed events had a type of `post-burst decrease' and resulted
from the eruption of a filament from an active region. An
impulsive non-thermal burst is typically observed before a
negative burst, because eruptions in strong magnetic fields of
active regions cause strong gyrosynchrotron emission. If the
eruption of a filament occurs out of active regions where magnetic
fields are weak, the enhancement of the microwave emission is
predominantly thermal and weak. In such a case, no pronounced
increase of the emission in total flux records is expected. If a
non-flare-related eruptive filament screens a compact radio
source, then an isolated negative burst is possible as was the
case in the 2011 December 13 event.

Two scenarios of screening have been revealed. In the first
scenario, which occurs in the majority of events, the shape and
structure of an eruptive filament persist. Such an expanding
filament moves away from the solar surface without dramatic losses
of mass and looks like a moving screen with increasing size and
decreasing opacity. Occasionally eruptions develop in anomalous
way. The eruptive filament essentially changes, while a part of
its material disperses over a large area and descends far from the
eruption site.

\subsection{Anomalous Eruptions}

The 1998 April 29 event with a huge darkening in an EIT 304~\AA\
image mismatching dimmings observed in coronal lines was presented
by \citet{ChertokGrechnev2003}. Initially this event was not
recognized as an anomalous eruption, which were unknown at that
time. A negative burst was recorded in Penticton. An anomalous
eruption in this event was revealed later \citep{Grechnev2011}.

The first well-studied anomalous eruption of a filament from an
active region occurred on 2004 July 13 \citep{Grechnev2008}. The
ejecta disintegrated into two parts, one of which flew away as a
decelerating CME, and another part returned to the Sun and
descended far from the eruption center. The latter part absorbed
background solar emission in various spectral ranges that was
observed as a negative microwave burst, widespread faint moving
dimmings at 195~\AA, and a huge dimming at 304~\AA, as
Figure~\ref{fig:20040713} shows in the middle row. An `EUV wave'
is visible in Figure~\ref{fig:20040713}f. The area of the
304~\AA\ darkening at a level of $-25\%$ was 6.7\% of the solar
disk.

NoRH 17 GHz images reveal a dark moving blob, which is outlined
with the white circle in Figure~\ref{fig:20040713} (top row). Its
lowest brightness temperatures reached 6000--8000~K against the
background of the quiet Sun (10\,000~K at 17~GHz) indicating its
large optical thickness and, consequently, the kinetic temperature
of $\sim 6000\,$~K. For comparison, the estimated average
temperatures of absorbers in Table~\ref{tab:burst_list} are
typically $\sim 10^4$~K. This blob probably showed a densest piece
of dispersed filament material.

The events of 2004 July 13 and 2011 June 7 appear to be similar.
On 2011 June 7, dark filament fragments were visible even in
original non-subtracted images in all coronal channels up to
94~\AA. Such strong absorption was observed for the first time.
\citet{Grechnev2011} contemplated probable properties of events
with anomalous eruptions. The 2011 June 7 event exhibited most of
these properties indeed. The event was accompanied by a powerful
flare, `EUV wave', and type II radio burst. The authors
anticipated higher probability for an anomalous course of an
eruption if it occurred in a complex magnetic configuration, which
was really the case in the 2011 June 7 event. By contrast, the
eruption on 2011 December 13 occurred outside of active regions,
and its course was not anomalous in accordance with expectations.

\subsection{A Possible Scenario for an Anomalous Eruption}

\citet{Gilbert2007} consider a few types of filament eruption
based on the traditional conception of a filament as a dense lower
part of a large-scale magnetic rope. Eruptions of filaments are
divided into three types: full eruptions, partial, and failed
ones. The type of an eruption is considered to be determined by
the position of the reconnection site relative to the filament
(below the filament, within its limits, or above it). If filament
material falls back on the Sun, then its returning trajectory is
considered to be the same as that of its lift-off. A possibility
of reconnection between magnetic structures of an eruptive
filament and surrounding magnetic fields in the corona was not
considered by the authors. However, decay of the magnetic
structure of an eruptive filament is possible in some events due
to such reconnection processes. The filament loses integrity in
such a situation.

\begin{figure*}
  \begin{center}
    \FigureFile(170mm,81mm){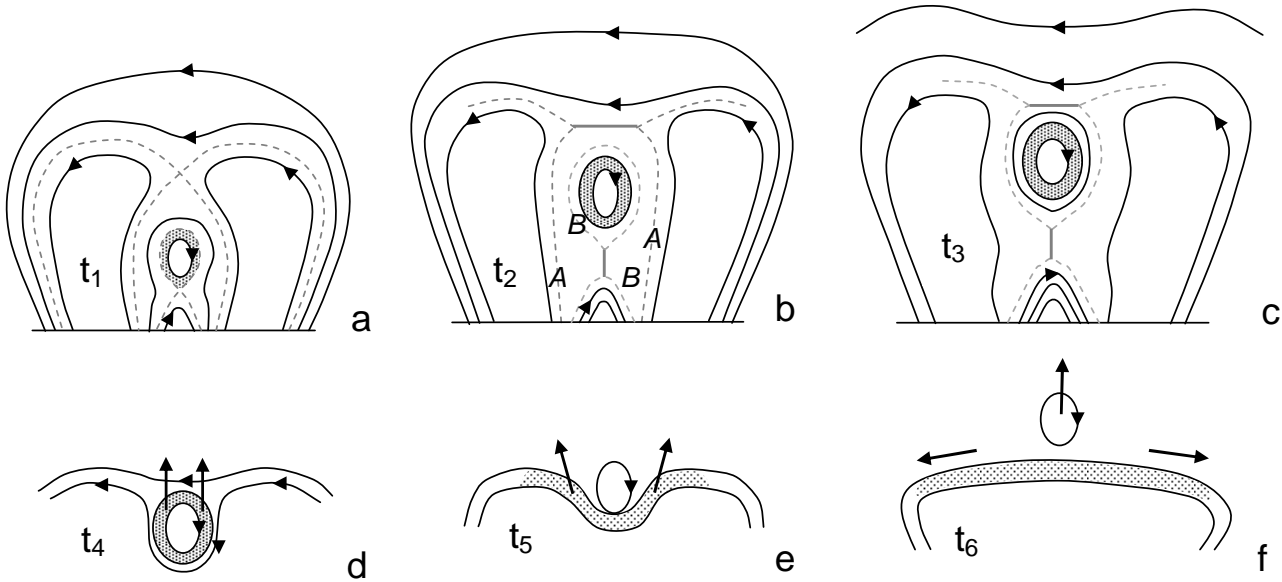}
  \end{center}
  \caption{A possible scenario of an anomalous eruption of an inverse
filament in a quadrupole configuration.}
 \label{fig:scenario_anomalous}
 \end{figure*}

Figure~\ref{fig:scenario_anomalous} presents a possible scenario
for an anomalous eruption of an inverse filament in a quadrupole
configuration. We consider an apex cross section of a
three-dimensional toroidal magnetic flux rope with ends rooted in
the photosphere. Such a toroidal flux rope is well visible in the
LASCO/C2 image of the 2011 June 7 CME in
Figure~\ref{fig:20110607_sdo_stereo}d. The initial filament in
Figure~\ref{fig:scenario_anomalous}a (time of $t_1$) is in the
equilibrium state. The dashed lines present the initial
separatrices. Let us follow evolution of plasma within the gray
ring cross section of the flux rope.

An upward rise of the flux rope leads to magnetic field
stretching, and a vertical current sheet develops. Standard flare
reconnection starts. With the flare onset ($t_2$ in
Figure~\ref{fig:scenario_anomalous}b), separatrices $AA$ and $BB$
approach. In the course of reconnection, the flux rope acquires
additional rings of the poloidal magnetic field, which creates a
progressively increasing propelling force. Acceleration of the
expanding flux rope increases. The flux rope exits into the outer
domain ($t_3$ in Figure~\ref{fig:scenario_anomalous}c). By this
moment, separatrices $AA$ and $BB$ have already merged. A
horizontal current sheet forms, and reconnection with the external
magnetic field starts to `undress' the flux rope thus making the
gray ring portion of interest naked. The poloidal flux starts to
decrease. Plasma of the filament is dispersed over the solar
surface (Figure~\ref{fig:scenario_anomalous}d--f). The
`undressing' of the gray ring portion and its heating goes on in
reconnection with external magnetic fields ($t_4-t_5$). The newly
formed magnetic tube straightens and becomes a static magnetic
loop. The plasma upflow ceases
(Figure~\ref{fig:scenario_anomalous}f). The sidewards velocity of
plasmas flowing from the top is almost equal to the upwards
velocity of the ejecta. The process results in dispersal over a
large area of material, which was previously confined within the
ring cross section.

\begin{figure*}
  \begin{center}
    \FigureFile(170mm,39mm){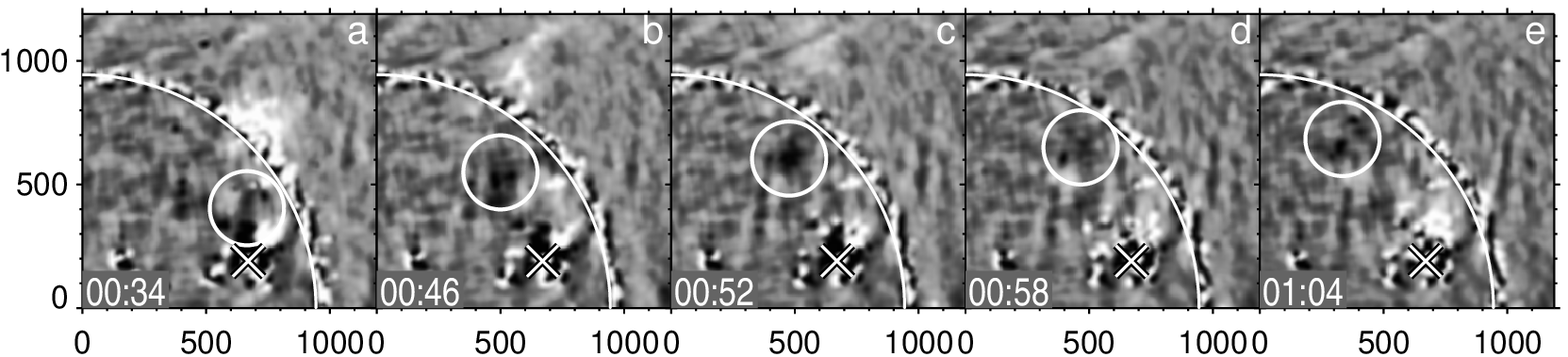}
    \FigureFile(170mm,44mm){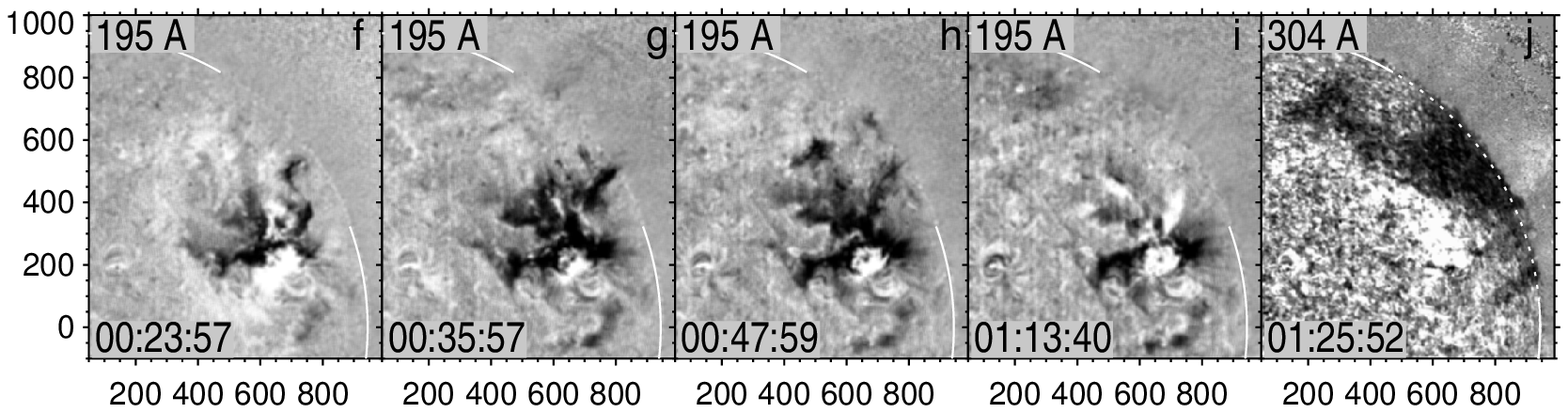}
    \FigureFile(170mm,44mm){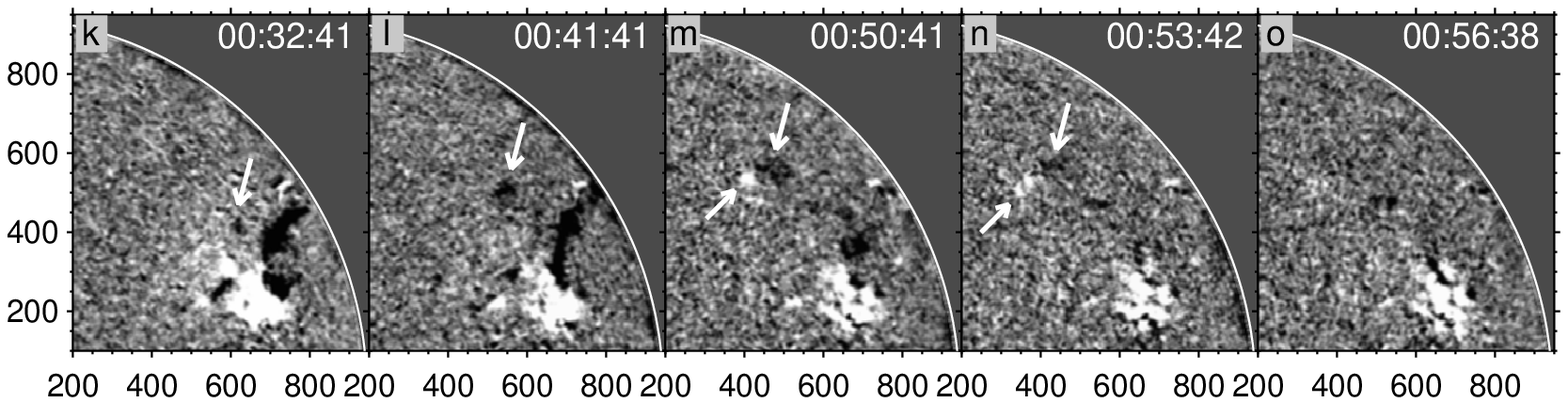}
  \end{center}
  \caption{Anomalous eruption on 2004 July 13.
a--e)~Moving dark absorbing feature (white circle) in NoRH 17 GHz
difference images. The slanted cross marks the eruption center.
f--j)~SOHO/EIT ratio images at 195~\AA\ and 304~\AA.
k--o)~H$\alpha$ ratio images (PICS, MLSO). The arrows indicate a
dark moving fragment of the disintegrated filament and remote
brightenings. The white arcs denote the solar limb.}
 \label{fig:20040713}
 \end{figure*}

\begin{figure*}
  \begin{center}
    \FigureFile(170mm,60mm){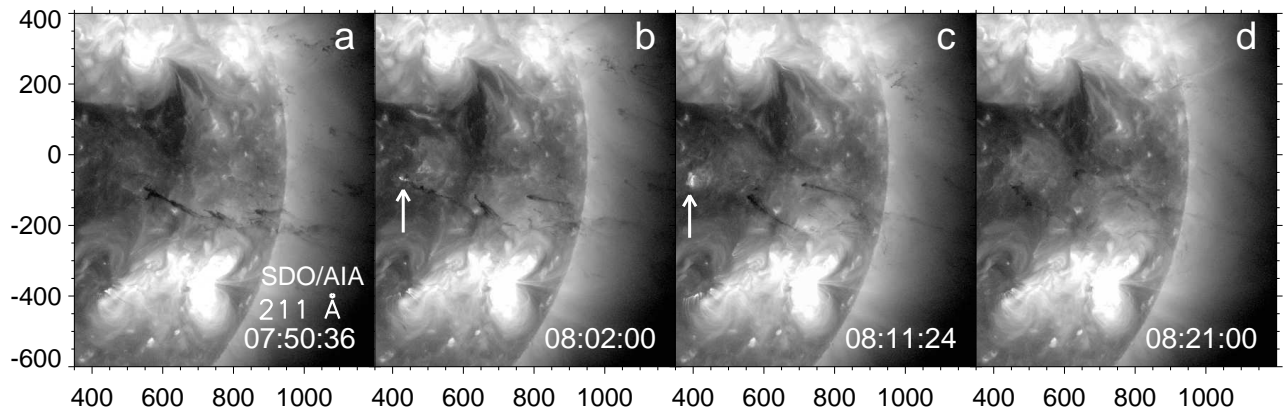}
  \end{center}
  \caption{SDO/AIA 211~\AA\ images of the 2011 June 7
anomalous eruption. The arrows indicate transient brightenings at
the places of falling filament fragments.}
 \label{fig:20110607_211}
 \end{figure*}

Note that blow of a fast and dense plasma downflow should produce
a shock in the transition region. Observations of two anomalous
eruptions confirm this expectation indeed.
Figure~\ref{fig:20040713}k--o (bottom row) shows H$\alpha$ ratio
images produced with the Polarimeter for Inner Coronal Studies
(PICS) of the Mauna Loa Solar Observatory (MLSO). Transient remote
brightenings produced by falling fragments of the former filament
are indicated by the arrows in these images.  Similar transient
brightenings obviously caused by downstreaming dark material are
visible in Figure~\ref{fig:20110607_211}, which presents SDO/AIA
211~\AA\ images (temperature sensitivity maximum 2~MK) of the 2011
June 7 event. These brightenings are visible in all coronal
channels of AIA up to 94~\AA\ and in the 1600~\AA\ channel. Such
remote non-active-region brightenings in the chromosphere were
interpreted many years ago by
\authorcite{Hyder1967a} (\yearcite{Hyder1967a,Hyder1967b}) as the
places of dissipation of kinetic energy of falling filament
material.

In the considered scenario, reconnection with external magnetic
field in the course of `undressing' of the flux rope tends to
increase its velocity due to upwards magnetic tension of
reconnected field lines (Figure~\ref{fig:scenario_anomalous}d,e),
while the decrease of the poloidal flux results in a decrease of
the total propelling force. The retarding influence of the
toroidal flux's tension and gravity persists. The scenario
predicts that the flux rope should sharply accelerate and then
gradually decelerate. This was really the case in the 2011 June 7
event. Similarly, all observed moving features only decelerated in
the 2004 July 13 event \citep{Grechnev2008}.

\section{Concluding Remarks}

Despite the rare occurrence of negative bursts, their studies have
provided new insight into solar eruptions. An important condition
of reaching success in these studies was a combined analysis of
microwave total flux and imaging data provided mainly by NoRP and
NoRH along with EUV images. These studies have revealed:

\begin{itemize}

 \item
 different kinds of negative bursts ranging from the well-known
post-burst decrease up to unusual isolated negative bursts without
preceding flare bursts;

  \item
 negative bursts caused by screening not only compact sources,
but also large quiet-Sun areas;

\item
 absorption of the background emission in material of an
eruptive filament, which can also be observed as a depression of
the He{\sc ii} 304~\AA\ line emission, usually without pronounced
counterparts in EUV emission lines;

 \item
 two scenarios of screening by either a steadily expanding filament
or remnants of a filament dispersed in an anomalous eruption;

 \item
 anomalous eruptions and their expected properties, which have been
inferred from scarce data of past years and were confirmed by
recent observations.

\end{itemize}

Most eruptions occur in the `normal' scenario: an eruptive
filament expands keeping its magnetic structure. The 2011 December
13 event considered here presents an interesting rare extremity of
an isolated negative burst.

Anomalous eruption of a filament with its disintegration appears
to evidence reconnection between magnetic fields of the
large-scale coronal environment and interior of the filament. In
such a case, the signs of the magnetic helicity must not coincide
for the ejected magnetic cloud and the active region from which
the eruption originated \citep{Uralov2013}. An anomalous eruption
presumably occurs if an eruptive filament passes through
vicinities of a coronal null point. One of possible scenarios of
an event with an anomalous eruption was proposed here.

As shown, parameters of ejected plasma projected onto the solar
disk can be estimated from multi-frequency total flux records of
negative bursts. Ongoing observations with NoRP and NoRH along
with EUV data can shed further light on different scenarios and
parameters of solar eruptions and can be used among the factors in
forecasting geoeffectiveness of Earth-directed magnetic clouds.

Being associated with eruptions on the Earth-facing solar surface,
microwave negative bursts indicate events, in which potentially
geoeffective CMEs develop. Especially hazardous can be anomalous
eruptions, whereas occurrence of negative bursts in such events is
not guaranteed. The most demonstrative example is the anomalous
eruption of 2003 November 18 responsible for the strongest
geomagnetic storm in last two decades \citep{Grechnev2013_I}, but
a negative burst could not be detected in this event due to
ongoing flaring. Similarly, the negative burst on 2003 May 28 was
interrupted by a subsequent flare. Microwave imaging observations
are very important also in this respect. Importance of anomalous
eruptions also calls for such requirements for future radio
telescopes as multi-frequency capability at long microwaves with a
field of view excessively covering the solar disk.

\bigskip

Acknowledgements. We thank K.~Shibasaki, H.~Nakajima, and
V.~Slemzin for fruitful discussions, V.~Zandanov and
S.~Anfinogentov for their assistance in data processing. We are
indebted to the anonymous referee for useful remarks. We are
grateful to instrumental teams operating Nobeyama solar
facilities, SSRT, EIT and LASCO on SOHO (ESA \& NASA), USAF RSTN
network, STEREO, and SDO missions.

This study was supported by the Russian Foundation of Basic
Research under grants 11-02-00757, 12-02-00037,
12-02-33110-mol-a-ved, and 12-02-31746-mol-a; the Integration
Project of RAS SD No.~4; the Program of basic research of the RAS
Presidium No.~22, and the Russian Ministry of Education and
Science under projects 8407 and 14.518.11.7047.



\end{document}